\documentclass[amsfonts,preprint,aps,nofootinbib,secnumarabic]{revtex4}
\usepackage{graphicx}

\def\lsim{\mathrel{\rlap{\lower4pt\hbox{\hskip1pt$\sim$}}}<}
\def\gsim{\mathrel{\rlap{\lower4pt\hbox{\hskip1pt$\sim$}}}>}
\begin{document}

\title{On: Natural Inflation}

\author{Katherine Freese}
\affiliation{Michigan Center for Theoretical Physics, Dept. of Physics,\\
  University of Michigan, Ann Arbor, MI 48109.\\
  Email: {\tt ktfreese@umich.edu}}

\author{William H. Kinney}
\affiliation{Dept.\ of Physics, Univerity at Buffalo, SUNY, Buffalo, NY 14260.\\
             Email: {\tt whkinney@buffalo.edu}}

\begin{abstract}
  We re-examine the model of Natural inflation, in which the inflaton
  potential is flat due to shift symmetries.  The original version of
  the model, where the inflaton is a pseudo Nambu-Goldstone boson with
  potential of the form $ V(\phi) = \Lambda^4 [1 \pm \cos(\phi/f)]$,
  is studied in light of recent data. We find that the model is alive
  and well. Successful inflation as well as data from the Wilkinson
  Microwave Anisotropy Probe require $f > 0.6 m_{\rm Pl}$ (where
  $m_{\rm Pl} = 1.22 \times 10^{19}$ GeV) and $\Lambda \sim m_{GUT}$,
  scales which can be accommodated in particle physics models. The
  detectability of tensor modes from natural inflation in upcoming
  microwave background experiments is discussed. We find that natural
  inflation predicts a tensor/scalar ratio within reach of future
  observations.
\end{abstract}

\pacs{98.80 Bp, Cq}

\maketitle

\section{Introduction}
The inflationary universe model was proposed \cite{guth} to solve
several cosmological puzzles, the horizon, flatness, and monopole
problems, via an early period of accelerated expansion.
To satisfy a combination of constraints on inflationary models, in
particular, sufficient inflation and microwave background anisotropy
measurements \cite{WMAP} of density fluctuations, the potential 
for the inflaton field must be very flat.  For
a general class of inflation models involving a single slowly-rolling
field (including new \cite{newinf}, chaotic \cite{chaotic}, and double
field inflation \cite{doublefield}, the ratio of the height to the
(${\rm width})^4$ of the potential must satisfy \cite{afg}
\begin{equation}
\label{eq:ratio}
\chi \equiv \Delta V/(\Delta \phi)^4 \le {\cal O}(10^{-6} - 10^{-8})
\, , 
\end{equation}
where $\Delta V$ is the change in the potential $V(\phi)$ and $\Delta
\phi$ is the change in the field $\phi$ during the slowly rolling
portion of the inflationary epoch.  Thus, the inflaton must be
extremely weakly self-coupled, with effective quartic self-coupling
constant $\lambda_{\phi} < {\cal O}(\chi)$ (in realistic models,
$\lambda_{\phi} < 10^{-12}$).  The small ratio of mass scales required
by Eq. (\ref{eq:ratio}) quantifies how flat the inflaton potential
must be and is known as the ``fine-tuning'' problem in inflation.
 
Three approaches have been taken toward this required flat potential
characterized by a small ratio of mass
scales.  First, some simply say that there are many as yet unexplained
hierarchies in physics, and inflation requires another one.  The hope
is that all these hierarchies will someday be explained.  In these
cases, the tiny coupling $\lambda_{\phi}$ is simply postulated {\it ad
  hoc} at tree level, and then must be fine-tuned to remain small in
the presence of radiative corrections.  But this merely replaces a
cosmological naturalness problem with unnatural particle physics.
Second, models have been attempted where the smallness of
$\lambda_{\phi}$ is protected by a symmetry, {\it e.g.,}
supersymmetry. In these cases (e.g., \cite{hrr}), $\lambda_{\phi}$ may
arise from a small ratio of mass scales; however, the required mass
hierarchy, while stable, is itself unexplained.  In addition, existing
models have limitations.  It would be preferable if such a hierarchy,
and thus inflation itself, arose dynamically in particle physics
models.

Hence, in 1990 we proposed a third approach, Natural Inflation
\cite{natural}, in which the inflaton potential is flat due to shift
symmetries.  Nambu-Goldstone bosons (NGB) arise whenever a
global symmetry is spontaneously broken.  Their potential is exactly
flat due to a shift symmetry under $\phi \rightarrow \phi + {\rm
  constant}$. As long as the shift symmetry is exact, the inflaton
cannot roll and drive inflation, and hence there must be additional
explicit symmetry breaking.  Then these particles become pseudo-Nambu
Goldstone bosons (PNGBs), with ``nearly'' flat potentials, exactly as
required by inflation.  The small ratio of mass scales required by
Eq. (\ref{eq:ratio}) can easily be accommodated. For example, in the
case of the QCD axion, this ratio is of order $10^{-64}$.  While
inflation clearly requires different mass scales than the axion, the
point is that the physics of PNGBs can easily accommodate the required
small numbers\footnote{For example, in
`invisible' axion models \cite{invisible} with Peccei-Quinn scale
$f_{PQ} \sim 10^{15}$ GeV, the axion self-coupling is $\lambda_a \sim
(\Lambda_{QCD}/f_{PQ})^4 \sim 10^{-64}$. (This simply reflects the
hierarchy between the QCD and GUT scales, which arises from the slow
logarithmic running of $\alpha_{QCD}$.)  Due to the nonlinearly
realized global symmetry, the potential for PNGBs is exactly flat at
tree level.  The symmetry may be explicitly broken by loop
corrections, as in schizon \cite{hillross} and axion \cite{axion}
models.  In the case of axions, for example, the PNGB mass arises from
non-perturbative gauge-field configurations (instantons) through the
chiral anomaly.  When the associated gauge group becomes strong at a
mass scale $\Lambda$, instanton effects give rise to a periodic
potential of height $\sim \Lambda^4$ for the PNGB field
\cite{instanton}.  Since the nonlinearly realized symmetry is restored
as $\Lambda \rightarrow 0$, the flatness of the PNGB potential is
natural in the sense of 't Hooft \cite{hooft}.}.

We first proposed this model and performed a simple analysis in
\cite{natural}. Then, in 1993, we followed with a second paper which
provides a much more detailed study \cite{natural2}.  The results of
Section III of the second paper, which presents a careful analysis of
the dynamics of the natural inflaton, are of particular relevance
here.  

Many types of candidates have subsequently been explored for natural
inflation.  For example \cite{kyy} used shift symmetries in Kahler
potentials to obtain a flat potential and drive natural chaotic
inflation in supergravity.  Additionally, \cite{hccr} examined natural
inflation in the context of extra dimensions and \cite{kapwei} used
PNGBs from little Higgs models to drive hybrid inflation.  Also,
\cite{firtye} and \cite{hsukallosh} use the natural inflation idea of
PNGBs in the context of braneworld scenarios to drive inflation.
Freese \cite{term} suggested using a PNGB as the rolling field in
double field inflation \cite{doublefield} (in which the inflaton is a
tunneling field whose nucleation rate is controlled by its coupling to
a rolling field).  We will focus in this paper on the original version
of natural inflation, in which there is a single rolling field; we
will comment further on other variants of natural inflation in Section
6.

In the current paper we show that the original proposal of natural
inflation is live and well, contrary to recent criticisms (which we
address in Section 3).  In particular, the single-field version of
the model is successful for $f > 0.6 m_{Pl}$ (and does not require $f
>> m_{Pl}$, contrary to the claims of \cite{hccr}).
A second focus of the current paper is to discuss tests
of natural inflation from existing and upcoming data from microwave
background experiments.  Recent results from the Wilkinson Microwave
Anisotropy Probe \cite{WMAP} are used to constrain our model, and
predictions are made for upcoming experiments such as the PLANCK
satellite which will begin taking data in 2007.

We begin in Section 2 by reviewing the basic idea of natural
inflation.  In Section 3 we present results of the evolution of the
scalar field driving inflation, including explicit numerical
calculation of the evolution of the scalar field in its potential.  In
Section 4 we discuss density fluctuations, and find the constraint on
the potential due to comparison with WMAP data. In Section 5 we
compute the tensor modes from natural inflation, and discuss their
detectability in upcoming microwave background experiments.  In
Section 6, we conclude with a discussion of the pros and cons of
having a model in which the width of the potential is of order the
Planck scale.

\section{Inflation due to Shift Symmetries}
Here we review the original variant of natural inflation
\cite{natural}, in which a single rolling field has a flat potential
due to a shift symmetry and drives inflation.  Whenever a global
symmetry is spontaneously broken, Nambu Goldstone bosons arise, with a
potential that is exactly flat due to a remaining shift symmetry under
$\phi \rightarrow \phi + {\rm constant}$.  If there is additional
explicit symmetry breaking, these particles become pseudo-Nambu
Goldstone bosons (PNGBs), with ``nearly'' flat potentials.  The
resulting PNGB potential in single field models (in four spacetime
dimensions) is generally of the form
\begin{equation}
\label{eq:potential}
 V(\phi) = \Lambda^4 [1 \pm \cos(N \phi/f)] \, .
\end{equation}
We will take the positive sign in Eq. (2) (this choice has no effect on
our results) and take $N = 1$, so the potential, of height $ 2
\Lambda^4$, has a unique minimum at $\phi = \pi f$ (we assume the
periodicity of $\phi$ is $2 \pi f$).

We show below that, for appropriately chosen values of the mass
scales, namely $f \sim m_{\rm Pl}$ and $\Lambda \sim m_{GUT} \sim
10^{15}$ GeV, the PNGB field $\phi$ can drive inflation.  This choice
of parameters indeed produces the small ratio of scale required by
Eq. (1), with $\chi \sim (\Lambda/f)^4 \sim 10^{-13}$.

We shall assume that inflation is initiated from a state that is at
least approximately thermal. In general, this is a dangerous assumption,
since there is no {\it a priori} reason to expect homogeneity or thermal
equilibrium prior to inflation. However, this assumption is in keeping
with the motivation of a PNGB potential arising from a phase transition
associated with spontaneous symmetry breaking. In addition, such a homogeneous,
thermal initial condition could naturally arise from an earlier period of
inflation associated with the breaking of the global symmetry at the
scale $f$.
For temperatures $T \leq f$, the global symmetry is spontaneously
broken, and the field $\phi$ describes the phase degree of freedom
around the bottom of a Mexican hat.  Since $\phi$ thermally decouples
at a temperature $T \sim f^2/m_{\rm Pl} \sim f$, we assume it is initially
laid down at random between 0 and $2\pi f$ in different causally
connected regions.  Within each Hubble volume, the evolution of the
field is described by
\begin{equation}
\label{eq:evolve}
\ddot\phi + 3H\dot\phi + \Gamma\dot\phi + V^\prime(\phi) = 0\ ,
\end{equation}
where $\Gamma$ is the decay width of the inflaton.  
In the temperature range $\Lambda \leq T \leq f$, the potential
$V(\phi)$ is dynamically irrelevant, because the forcing term
$V^\prime(\phi)$ is negligible compared to the Hubble-damping term.
(In addition, for axion models, $\Lambda \rightarrow 0$ as 
$T/\Lambda \rightarrow \infty$ due to the high-temperature suppression
of instantons \cite{instanton}.)
Thus, in this temperature range, aside from the
smoothing of spatial gradients in $\phi$, the field does not evolve.
Finally, at $T \leq \Lambda$, in regions of the universe with $\phi$
initially near the top of the potential, the field starts to roll
slowly down the hill toward the minimum.  In those regions, the energy
density of the universe is quickly dominated by the vacuum
contribution $(V(\phi) \simeq 2\Lambda^4 \geq \rho_{rad} \sim T^4)$,
and the universe expands exponentially. 
 
To successfully solve the cosmological puzzles of the standard
cosmology, an inflationary model must satisfy a variety of
constraints.  We describe these constraints in the following sections.

\section{Evolution of the Inflaton Field}

In this section we present results for the evolution of the scalar
field driving natural inflation.  First, we review the standard slow
roll (SR) analysis, and then turn to the results of an exact
calculation obtained by numerically solving the equations of motion.
As our result we find that sufficient inflation takes place as long as
\begin{equation}
f > 0.06\,\, m_{\rm Pl}.
\end{equation}
In Section 4 we will derive stronger bounds on $f$
from constraints on the spectral index of density fluctuations.
Throughout, we take $m_{\rm Pl} = 1.22 \times 10^{19}$ GeV.
Hereafter, we take the onset of inflation to take place at a field
value $0<\phi_1/f <\pi$, and the end of inflation to be at a field
value $0<\phi_2/f<\pi$.

\subsection{Standard slow roll analysis}

A sufficient, but not necessary, condition for inflation is that the
field be slowly rolling, {\it i.e.} its motion is overdamped, $\ddot\phi <<
3H\dot\phi$. The SR condition implies that two conditions are met:
\begin{equation}
\label{eq:sr}
\left|V^{\prime\prime}(\phi)\right|
< 9H^2\ ,\ {\rm i.e.,}\quad  \sqrt{{2\left|\cos(\phi/f)\right|\over 1+
\cos(\phi/f)}} < {\sqrt{48\pi}f \over m_{\rm Pl}}
\end{equation}
and
\begin{equation}
\label{eq:sr2}
\left|{V^\prime(\phi) m_{\rm Pl} \over
V(\phi)}\right| < \sqrt{48\pi}\ ,\ {\rm i.e.,}\quad {\sin(\phi/f) \over
1+ \cos(\phi/f)} < {\sqrt{48\pi} f \over m_{\rm Pl}}\ .
\end{equation}
From Eqs.. (\ref{eq:sr}) and (\ref{eq:sr2}), the existence of a broad
SR regime requires $f \geq m_{\rm Pl} \big/ \sqrt{48\pi}$ 
and ends when $\phi$ reaches a
value $\phi_2$, at which one of the inequalities (\ref{eq:sr}) or
(\ref{eq:sr2}) is violated.  For example, for $f = m_{\rm Pl}$,
$\phi_2 /f = 2.98$ (near the minimum of the potential), while for $f =
m_{\rm Pl} \big/ \sqrt{24\pi}$, $\phi_2 /f = 1.9$. Clearly, as $f$
grows, $\phi_2/f$ approaches $\pi$.
We note that the conditions (\ref{eq:sr}) and (\ref{eq:sr2}) are
approximate relations.  A more precise calculation using the slow roll
parameters $\epsilon$ and $\eta$ gives similar bounds.  
Next we present exact numerical solutions of the equations of motion to
substantiate our results. 
\subsection{Numerical Evolution of the Scalar Field}

In \cite{natural2}, we obtained exact numerical solutions to the
equations of motion for the inflaton in the natural inflation model.
We briefly recapitulate results from a numerical evolution of the
scalar field found in Section III of \cite{natural2}, which provides
more precise results than the simple SR analysis.  Without loss of
generality, we take the initial velocity \footnote{Nonzero initial
  velocities have been studied numerically by \cite{knoxolinto}. They
  find that, due to the periodic nature of the potential, the effect
  of initial velocities is merely to shift, but not change the size
  of, the phase space of initial field values which lead to at least
  60 e-folds of inflation.}  to be $v_1 \equiv \dot \phi_1/Hf =0$.

We find that the exact solution roughly reproduces the results of the
SR analysis presented previously. As long as $f>0.1 m_{\rm Pl}$, the
results agree to within 10\%.  In particular, the numerical results
for the maximum field value at the start of inflation, $\phi^{max}_1$,
are nearly identical to the SR estimates for values of $f$ near
$m_{\rm Pl}$; they differ by $\sim 10\%$ for $f = m_{\rm Pl}$ and
deviate significantly as $f$ approaches $m_{\rm Pl}/\sqrt{24 \pi}$
from above.  Further details can be found in \cite{natural2}.

\subsection{Sufficient inflation.}

The expansion $H=\dot a/a$ of the universe is determined
by the Friedmann equation,
\begin{equation}
\label{eq:fried}
H^2 = {8 \pi \over 3 m_{\rm Pl}^2} \left[ V(\phi) 
+ {1 \over 2} \dot\phi^2\right] .
\end{equation}
Inflationary expansion takes place when the potential $V$ dominates in
the energy density.  To solve the flatness and horizon problems,
we demand that the scale factor of the universe
inflates by at least 60 e-foldings during the SR regime,
\begin{equation}
\label{eq:suffice}
N_e(\phi_1,\phi_2,f) \equiv \ln(R_2/R_1) =
\int^{t_2}_{t_1} Hdt = {-8\pi \over {m_{\rm Pl}}^2} \int_{\phi_1}^{\phi_2}
{V(\phi) \over V^\prime(\phi)} d\phi
= {16\pi f^2 \over {m_{\rm Pl}}^2} \ln\left[{\sin(\phi_2/2f)
\over \sin(\phi_1/2f)}\right] \geq 60\ .
\end{equation}
Using Eqs. (\ref{eq:sr}) and (\ref{eq:sr2}) to determine $\phi_2$ as
a function of $f$, the constraint (\ref{eq:suffice}) determines the
maximum value ($\phi^{max}_1$) of $\phi_1$ consistent with sufficient
inflation.  The fraction of the universe with
$\phi_1\in[0,\phi^{max}_1]$ will inflate sufficiently.  The
requirement that a reasonable fraction of the universe inflate
sufficiently places a bound on $f$.

There are two conceptually different approaches to the question of
what fraction of the universe inflates sufficiently, and hence to the
bound on the scale $f$.  The first is an ``a priori probability.'' In
this (more restrictive) approach, one determines the fraction of the
volume of the universe {\it before} inflation which will inflate
sufficiently, and requires this fraction to be reasonably large.  If
we assume that $\phi_1$ is randomly distributed between 0 and $\pi f$
from one horizon volume to another, the probability of being in a
region of the universe that inflates enough is $\phi^{max}_1 / \pi f$.
For example, for $f = 3 m_{\rm Pl}$, $m_{\rm Pl}$, $m_{\rm Pl}/2$, and
$m_{\rm Pl} / \sqrt{24\pi}$, the probability is 0.7, 0.2, $3 \times
10^{-3}$, and $3\times 10^{-41}$.  The fraction of the universe that
inflates sufficiently drops precipitously with decreasing $f$, and
hence restricts $f$ to be near $m_{\rm Pl}$.  However, this approach
is unnecessarily restrictive.

The second approach, namely ``a posteriori probability'', is more
sensible. Here one examines the universe after inflation has taken
place, and ascertains what fraction of the {\it final} volume of the
universe has inflated sufficiently to look like our own.  After
inflation, those initial Hubble volumes of the universe that did
inflate end up occupying a {\it much} larger volume than those that
did not.  This second approach is much less restrictive and allows a
lower value of $f$, as shown below. We note that neither of these
approaches addresses the broader (and unsolved) issue of how to
rigorously define a measure on the space of initial conditions for
inflation, since we are implicitly assuming homogeneity and thermal
equilibrium. However, these arguments do serve to establish the
plausibility and naturalness of the model.

\subsection{A Posteriori Probability of Sufficient Inflation}
 
We now calculate the {\it a posteriori} probability of sufficient
inflation.  We consider the universe at the end of inflation, and
calculate the fraction $P$ of the volume of the universe at that time
which had inflated by at least 60 e-foldings:
\begin{equation}
\label{eq:apost}
P = 1 - { \int_{\phi_1^{max} }^{\pi f} \, d\phi_1 \exp[3 N(\phi_1)] \over
\int_{H/2\pi}^{\pi f} \, d\phi_1 \exp[3 N(\phi_1)] } . 
\end{equation}
Here, the lower limit of integration in the denominator is the limit
of validity of the semiclassical treatment of the scalar field; the
initial value of $\phi$ must exceed its quantum fluctuations, $\phi_1
\geq \Delta \phi = H/2\pi$.  This fraction $P$ is then the {\it a
  posteriori} probability of sufficient inflation.

Our basic result \cite{natural2} is that the {\it a posteriori}
probability for inflation $P$ is essentially unity for $f$ larger than
the critical value $f_c \simeq 0.06 m_{\rm Pl}$.  As $f$ drops below
this value, the probability given by Eq. (\ref{eq:apost}) rapidly
approaches 0.  Hence the requirement that a significant fraction of
the universe inflate sufficiently places a lower bound on the scale
\begin{equation}
\label{eq:boundsuff}
f > f_c \simeq 0.06 m_{\rm Pl} .
\end{equation}  

We have explicitly calculated the evolution of the scalar field in
natural inflation and found that the claim of \cite{hccr} that
$f>>m_{\rm Pl}$ is unnecessarily restrictive. The correct bound
due to sufficient inflation is given by Eq. (\ref{eq:boundsuff}).

\section{Density Fluctuations}
 
The amplitude and spectrum of density fluctuations produced in the
natural inflation model can be compared with microwave background data
in order to constrain the height and width of the potential.  Here we
find the constraint on the potential due to comparison with WMAP data.
 
Quantum fluctuations of the inflaton field as
it rolls down its potential generate adiabatic density perturbations
that may lay the groundwork for large-scale structure and
leave their imprint on the microwave background anisotropy 
\cite{pert1,pert2,pert3}.
In this context, a convenient measure of the perturbation
amplitude is given by the gauge-invariant variable $\zeta$, first
studied in \cite{bst}.  We follow \cite{sbb}
in defining the power in $\zeta$,
\begin{equation}
\label{eq:densfluc}
P^{1/2}_\zeta(k) = {15\over 2}\left({\delta \rho \over \rho}\right)_{HOR}
=  {3\over 2\pi}{H^2 \over \dot \phi}. 
\end{equation}
Here, $(\delta \rho/\rho)_{HOR}$ denotes the perturbation amplitude
(in uniform Hubble constant gauge) when a given wavelength enters the
Hubble radius in the radiation- or matter-dominated era, and the last
expression is to be evaluated when the same comoving wavelength
crosses outside the Hubble radius during inflation.  For
scale-invariant perturbations, the amplitude at Hubble-radius-crossing
is independent of perturbation wavelength.  Normalizing to the COBE
\cite{cobe} or WMAP \cite{WMAP} Cosmic Microwave Background (CMB) 
anisotropy measurements gives
$P^{1/2}_\zeta(k) \sim 10^{-5}$.  We can use this normalization to get
an approximate fix on the scale $\Lambda$. Using the analytic
estimates of Sec. IIIA, the largest amplitude perturbations on
observable scales are produced 60 e-foldings before the end of
inflation, where $\phi = \phi^{max}_1$, and have amplitude
\begin{equation}
\label{eq:319}
P^{1/2}_{\zeta}
\simeq {\Lambda^2 f \over m_{\rm Pl}^3}{9\over 2\pi}\left({8\pi
\over 3}\right)^{3/2} {[1+\cos(\phi^{max}_1/f)]^{3/2} \over
\sin(\phi^{max}_1/f)} \, .
\end{equation}
We can obtain an analytic estimate of $\Lambda$ as a function of $f$
when $f \le (3/4)m_{\rm Pl}$; in this case, it is a
good approximation to take $\phi^{max}_1/\pi f \ll 1$. As a result,
in Eq. (\ref{eq:319}),  we have approximately
\begin{equation}
\label{eq:321}
P^{1/2}_{\zeta}  \approx {1.4 \Lambda^2 f \over M^3_{\rm Pl}}
\left({16 \pi \over 3}\right)^{3/2} \left({f\over \phi^{max}_1}\right)
~ . 
\end{equation}
Now the last term in this expression is obtained by using
Eq. (\ref{eq:suffice}) with $N(\phi^{max}_1,\phi_2,f) = 60$:
\begin{equation}
\label{eq:322}
{\phi^{max}_1 \over f} \simeq 2 \sin \left({\phi_2 \over 2f}\right)
{\rm exp} \left[ - {15 m^2_{\rm Pl} \over 4 \pi f^2}\right]
~ . 
\end{equation}
Applying the CMB normalization constraint to Eq. (\ref{eq:319}) gives 
$\Lambda \sim 10^{15}{\rm GeV}-10^{16}{\rm GeV}$ for $f \sim m_{\rm Pl}$.
Thus, to generate the fluctuations responsible for large-scale structure,
$\Lambda$ should be comparable to the GUT scale, and the inflaton mass
$m_{\phi} = \Lambda^2/f \sim (10^{11}-10^{13})$ GeV.
We note that this is strictly only an upper bound on the scale
$\Lambda$, since the perturbations responsible for large-scale
structure could be formed by some other (non-inflationary) mechanism.
 
\subsection{Density Fluctuation Spectrum}
 
Using the approximations above, we can investigate the wavelength
dependence of the perturbation amplitude at Hubble-radius-crossing and
in particular study how it deviates from scale-invariance (usually
associated with inflation).
 
Let $k$ denote the comoving wavenumber of a fluctuation.  The comoving
length scale of the fluctuation, $k^{-1}$, crosses outside the comoving
Hubble radius $[Ha]^{-1}$ during inflation at the time when the
rolling scalar field has the value $\phi_k$.  This occurs $N_I(k)
\equiv N(\phi_k,\phi_2,f)$ e-folds before the end of inflation, where
$N(\phi_k,\phi_2,f)$ is given by Eq. (\ref{eq:suffice}) 
with $\phi_1$ replaced by
$\phi_k$. The corresponding comoving length scale (expressed in current
units) is
\begin{equation}
\label{eq:324}
k^{-1} \simeq (3000 h^{-1} {\rm Mpc}) \exp(N_I(k)-60) ~ , 
\end{equation}
where the horizon size today is $\simeq 3000 h^{-1} {\rm Mpc}$.
For scales of physical interest for large-scale structure, $N_I(k) \ge 50$;
for $f \le (3/4)m_{\rm Pl}$, these scales satisfy $\phi_k/f \ll 1$.
In this limit, comparing two different field values $\phi_{k_1}$
and $\phi_{k_2}$, from Eq. (\ref{eq:suffice}) we have
\begin{equation}
\label{eq:325}
\phi_{k_2} \simeq \phi_{k_1} \exp \left(-~{\Delta N_I m^2_{\rm Pl}\over 16 \pi
f^2}\right) ~ , 
\end{equation}
where $\Delta N_I = N_I(k_2) - N_I(k_1)$. Thus, using Eqs. (\ref{eq:319}) and (\ref{eq:321}),
we can compare the perturbation amplitude at the two field values,
\begin{equation}
\label{eq:326}
{(P^{1/2}_{\zeta})_{k_1} \over (P^{1/2}_{\zeta})_{k_2}} \simeq
{\phi_{k_2}\over
\phi_{k_1}} \simeq \exp \left(-~{\Delta N_I m^2_{\rm Pl}\over 16 \pi
f^2}\right) ~ . 
\end{equation}
Now, from Eqs. (\ref{eq:324}), we have the relation $\Delta N_I =
\ln({k_1}/{k_2})$
(here we have approximated $H_{k_1} \simeq H_{k_2}$;
more precisely, $\Delta N_I = \ln({k_1 H_{k_2}}/{k_2 H_{k_1}})$).
Substituting this relation into (\ref{eq:326}), we find how the
perturbation amplitude at Hubble radius
crossing scales with comoving wavelength,
\begin{equation}
\label{eq:327}
\left({\delta \rho \over \rho}\right)_{HOR,k} \sim
(P^{1/2}_{\zeta})_{k}\sim k^{-m^2_{\rm Pl}/16 \pi f^2} .
\end{equation}
By comparison, for
a scale-invariant spectrum, the Hubble radius amplitude would
be independent of the perturbation length scale $k^{-1}$; the positive
exponent in Eq. (\ref{eq:327}) indicates that the PNGB models
with $f \le m_{\rm Pl}$ have more
relative power on large scales than do scale-invariant fluctuations.
 
It is useful to transcribe this result in terms of the power
spectrum of the primordial
perturbations at fixed time (rather than at
Hubble-radius crossing). Defining the Fourier transform $\delta_k$
of the density field, from Eq. (\ref{eq:327})
the power spectrum is a power law in the
wavenumber $k$,
${|\delta_k|^2 }\sim
k^{n_s}$, where the index $n_s$ is given by
\begin{equation}
\label{eq:deviate}
n_s = 1 - {m^2_{\rm Pl}\over 8 \pi f^2} ~~~~(f \le 3m_{\rm Pl}/4) ~ .
\end{equation}
For comparison, the scale-invariant
Harrison-Zel'dovich-Peebles-Yu spectrum corresponds to $n_s = 1$.
For values of $f$ close to $m_{\rm Pl}$, the spectrum
is close to scale-invariant, as expected; however, as $f$ decreases,
the spectrum deviates significantly from scale-invariance--{\it e.g.,}
for $f = m_{\rm Pl}/\sqrt{8\pi} = 0.2 m_{\rm Pl}$, the perturbations
have a white noise spectrum, $n_s = 0$. 

Recently WMAP has placed bounds on the spectrum of density
fluctuations.  If we assume that inflationary perturbations
are indeed responsible for what is being seen in the WMAP data,
then these spectral bounds can be translated into bounds
on the parameter $f$ in the potential.  The precise formulation
of the WMAP results depends on the choice of priors.  Here we take the 
bound on the deviation of the spectrum from scale invariant from WMAP
as found by \cite{will2, barger}:
\begin{equation}
\label{eq:wmap}
|n_s - 1| < 0.1 \, .
\end{equation}
Applying this bound to Eq. (\ref{eq:deviate}),
we see that a strong lower bound on the scale $f$
results:
\begin{equation}
\label{eq:wmapbound}
f \geq 0.6 m_{\rm Pl} .
\end{equation}
This is the strongest bound on the scale $f$.

\section{Tensor Modes}

In addition to density fluctuation, inflation also predicts the generation
of tensor (gravitational wave) fluctuations with amplitude
\begin{equation}
P_{\rm T}^{1/2} = {H \over 2 \pi}.
\end{equation}
In this section we study these tensor modes and discuss their
detectability in upcoming microwave background experiments. We also
examine the possible running of the scalar index 
and find that it is so small as to be observationally inaccessible.

For comparison with observation, the tensor amplitude is conventionally
expressed in terms of the tensor/scalar ratio $r$, defined as \footnote{
Normalization of this parameter varies in the literature. We use the
convention of Peiris {\it et al.}\cite{peirisetal}}
\begin{equation}
r \equiv {P_{\rm T}^{1/2} \over P^{1/2}_{\zeta}} = 16 \epsilon,
\end{equation}
where $\epsilon$ is the first slow roll parameter evaluated when the
fluctuation mode crosses the horizon, $\phi = \phi_1^{max}$:
\begin{eqnarray}
\epsilon &=& {m_{\rm Pl}^2 \over 16 \pi^2} \left(V'(\phi_1^{max}) \over V(\phi_1^{max})\right)^2\cr
&=& {1 \over 16 \pi^2} \left({m_{\rm Pl} \over f}\right)^2 \left[{\sin{(\phi_1^{max} / f)} \over 1 + \cos{(\phi_1^{max} /f)}}\right]^2\cr
&\simeq& {1 \over 32 \pi^2} \left({m_{\rm Pl} \over f}\right)^2 \left({\phi_1^{max} \over f}\right)^2,\ \phi \ll f.
\end{eqnarray}

In principle there are four parameters describing the scalar and
tensor fluctuations: the amplitudes and spectra of both components.
The amplitude of the scalar perturbations is normalized by the height
of the potential (the energy density $\Lambda^4$).The tensor spectral
index $n_T$ is not an independent parameter since it is related to the
tensor/scalar ratio by the inflationary consistency condition $r = -8
n_{\rm T}$.  The remaining free parameters are the spectral index $n$
of the scalar density fluctuations, and the tensor amplitude (given by
$r$).  

Hence, a useful parameter space for plotting the model predictions
versus observational constraints is on the $(r,n)$ plane
\cite{kinney98,dodelson97}.  Natural inflation generically predicts a
tensor amplitude well below the detection sensitivity of current
measurements such as WMAP. However, the situation will improve
markedly in future experiments with greater sensitivity such as the
Planck satellite, which will start taking data in 2007, and proposed
experiments such as CMBPOL. 

Figure 1 shows the predictions of Natural inflation for various
choices of the number of e-folds $N_I$ and the mass scale $f$,
together with a variety of observational constraints.  Fluctuations on
observable scales (up to the scale of the current horizon size) are
expected to lie roughly in the range $N_I = 50 - 60$, depending on the
reheat temperature (although the relevant range also depends on the
subsequent evolution of the universe \cite{liddle2003}).  In general,
a lower value of $f$ results in a ``redder'' (smaller $n$) spectrum
and a smaller tensor fluctuation amplitude.  The current observational
constraint from WMAP is given by the shaded (green) region on the left
hand of the plot: the white region is still allowed by WMAP.  We have
also forecast error bars for the PLANCK satellite based on a Fisher
matrix analysis (see Ref.  \cite{kinney98} for details of the
calculation).  Roughly, the PLANCK satellite is expected to have
1$\sigma$ error bars $\sim \pm 0.05$ on the magnitude of $r$, and
1$\sigma$ errors bars $\sim \pm 0.01$ on $n$.  The hatched (blue)
region indicates the 2$\sigma$ sensitivity of the PLANCK satellite if
the central value is (arbitrarily) chosen to be $r \sim 0.01$.  The
central value is arbitrary; only the size of the error bars is
significant. Similarly, we have also forecast error bars for a
hypothetical experimental measurement with the same angular resolution
as Planck, but with sensitivity improved by a factor of three; the
solid (black) error ellipse is the corresponding result (the $1
\sigma$ errors on $r$ here are roughly $\pm 5 \times 10^{-3}$).  Hence
PLANCK should be able to detect the tensor signal from natural
inflation if $f > 1.5 m_{Pl}$.  The next generation of experiments
should be able to do so for $f > 0.7\ m_{\rm Pl}$, the region allowed
by WMAP data.

One property of the potential to note is that the spectral index is
very weakly dependent on $N_I$ for $f < m_{\rm Pl}$, indicating that
the ``running'' of the spectral index $d n / d \ln{k}$ is negligible.
A more careful calculation indicates that the running of the spectral
index is less than $10^{-3}$ for all parameter regions considered
here, and therefore for all practical purposes unobservable. This
provides a powerful means of falsifying Natural Inflation. In particular, if
indications of a strong negative running of the spectral index 
\cite{Rebolo:2004vp} from small-scale CMB observations such as CBI 
\cite{Mason:2002tm}, ACBAR \cite{Kuo:2002ua} and VSA 
\cite{Dickinson:2004yr} are borne out, this will kill the model, 
at least in its simplest single-field form of Eq. (\ref{eq:potential}).

\begin{figure}
\centerline{\includegraphics[width=5.5in]{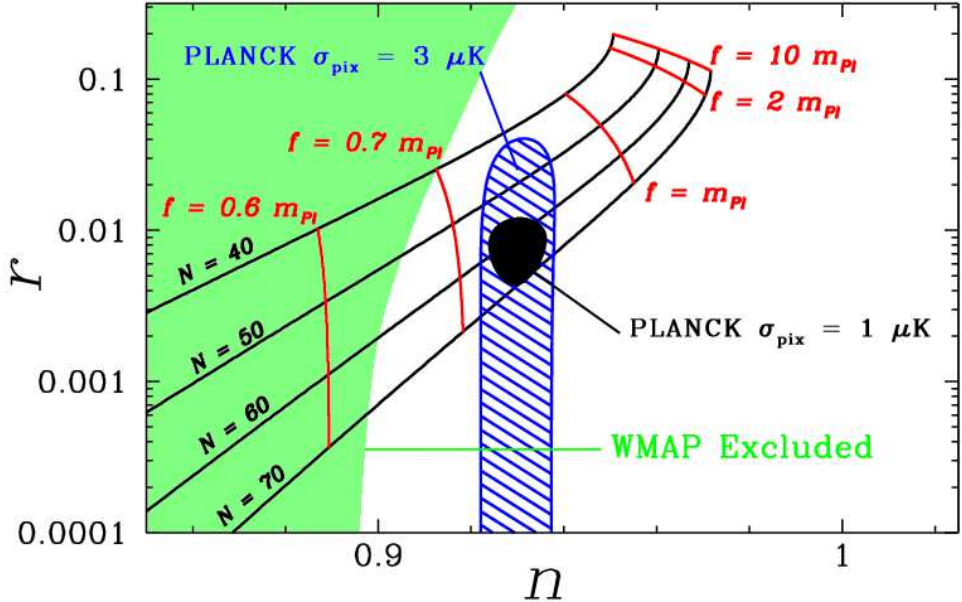}}
\caption{The predictions of Natural Inflation compared with 
  current and projected observational constraints, plotted on the
  $(r,n_s)$ plane, where $r$ is the tensor/scalar ratio and $n_s$ is
  the spectral index of scalar fluctuations. The lines show the
  predictions of natural inflation for varying choices of the mass
  scale $f$ and the number of e-folds $N_I$. Length scales of order
  the current horizon size correspond to $N_I \simeq 60$ for high
  reheat temperature. In general, a lower value of $f$ results in a
  ``redder'' (smaller $n_s$) spectrum and a smaller tensor fluctuation
  amplitude. The shaded region at the left of the plot (green) is
  excluded to $2\sigma$ by WMAP \cite{will2}. The hatched (blue) error
  ellipse is the $2 \sigma$ sensitivity expected for the Planck
  satellite. The central value is arbitrary: only the size of the
  error bar is significant.  The solid (black) error ellipse is the
  corresponding result for a hypothetical experiment with the same
  angular resolution as Planck but with a factor of three better
  temperature sensitivity. Such a measurement would be capable of
  detecting the gravitational wave fluctuations from Natural
  Inflation.}
\end{figure}

\section{Discussion}

In conclusion, natural inflation is alive and well.  Recent WMAP data
constrain the width of the potential to be $f >0.6 m_{\rm Pl}$, and
our predictions show that upcoming CMB observations such as the PLANCK
satellite may be able to see the tensor modes.

In this section, we discuss the pros and cons of $f \sim m_{\rm Pl}$,
as well as comment on some of the literature of PNGB models using
shift symmetries.

Although it is {\it not} true that the original model of natural
inflation requires $f>>m_{\rm Pl}$ for the width of the potential, it
does require $f$ to be of order $m_{\rm Pl} \sim 10^{19}$ GeV.  In
fact, virtually all 4D inflationary models require $f\sim m_{\rm Pl}$
and the height of the potential $\sim m_{GUT}$, in order to satisfy
the simultaneous requirements of sufficient inflation and the right
amplitude of density perturbations; this fact is emphasized by the
conclusions of Ref. \cite{afg}. However, the height of the potential
is generically of order the GUT scale, far enough below the Planck
scale that we can safely ignore quantum gravitational effects on the
background evolution. However, energy density is not the only issue.

In \cite{kammarch, holman}, it was argued that Planck-scale physics
results in the violation of all global symmetries, including the
Peccei Quinn symmetry of the axion and the underlying symmetry from
which we derive the PNGB inflaton.  Wormholes are suggested as one
mechanism for this violation, and black holes another (as a
consequence of black hole no-hair theorems, the global charge of a
black hole is not defined). The authors argue that, as a consequence,
one is required to add all higher dimension operators (suppressed by
powers of $m_{\rm Pl}$) consistent with the symmetries of the full
theory, which then does not respect global symmetries.  One should
include terms of the form:
\begin{equation}
{\cal L} = {1 \over 2} m^2 \phi^2 + \lambda \phi^4 
+ \sum_{n = 6}^{\infty}{\lambda_n \left(\phi^n \over m_{\rm Pl}^{n-4}\right)}.
\end{equation}
Without a complete theory of quantum gravity, the validity of these
arguments is not clear.  If true, then the idea of using a PNGB
directly as the inflaton would fail; however, the axion also could not
exist and we would have no theory at all to escape the strong CP
problem in QCD.

In \cite{lyth}, Lyth discussed the failure of effective field theory
if the width of an inflationary potential approaches $m_{\rm Pl}$.
Again, if inflation is to be formulated as an effective low-energy
field theory, he argues that we expect additional nonrenormalizable
operators in the Lagrangian to be suppressed by inverse powers of $m_{\rm
  Pl}$ as above.  Then if observational constraints require the field
to travel a distance $\Delta \phi \sim m_{\rm Pl}$, the effective
field theory will begin to break down due to radiative corrections
from the nonrenormalizable operators. Such a theory rapidly becomes
inconsistent as $\Delta \phi \gg m_{\rm Pl}$.  Motivated by the desire
to evade these issues, in 1995 Kinney and Mahantappa \cite{km}
constructed natural inflation models in which symmetries suppress the
mass terms and the potential is of the form $V \simeq 1 - \sin^4{(\phi
  / f)} \sim 1 - \phi^4$.  Then one automatically obtains an effective
width of the potential $f << m_{\rm Pl}$ \cite{knox92}.
One does so, however,
at the expense of an unobservably small tensor component in the CMB.

Natural inflation has been implemented in the context of extra
dimensions with varying degrees of success.  Recently, Arkani-Hamed
{\it et al.} \cite{hccr} examined natural inflation in the context of
extra dimensions, and also found models for which the mass terms were
suppressed by a symmetry with $f << m_{\rm Pl}$, similar to the work
of \cite{km}.  Focusing on a Wilson line in a fifth
dimension, Arkani-Hamed {\it et al.} alternatively suggested that one
might obtain large $f >> m_{\rm Pl}$ if the inflaton is the extra
component of a gauge field propagating in the bulk.

However, Banks {\it et al.} examined the general question of whether
it is ever possible to obtain large values of $f>> m_{\rm Pl}$ in
string theory \cite{Banks:2003sx}.  While their study was not
exhaustive, it strongly suggests that it is not possible.  Generically
there is a T-dual description in which small radii become large and
the value of $f$ is small ($f<< m_{\rm Pl}$); hence the model of
Arkani-Hamed {\it et al.} does not succeed in providing large $f$
inflation.  In addition, in a variety of regions in moduli space Banks
{\it et al.} find that there are low action instantons which give rise
to rapidly varying contributions to the potential that effectively
rescale the value of $f$ to the Planck scale.  Although they do not
provide an exhaustive proof, Banks {\it et al.} suggest the very
strong statement that natural inflation cannot work in the context of
string theory for $f>>m_{\rm Pl}$.  However, natural inflation with $f
\sim m_{\rm Pl}$, as discussed throughout this paper, is fine.

Natural inflation has been implemented in brane-world scenarios as
well.  Shift symmetries have been studied in brane inflation by
Firouzjahi and Tye \cite{firtye} and in the work of Hsu and Kallosh
\cite{hsukallosh}. The four dimensional effective field theory
description of some braneworld scenarios is likely to be described by
the physics in this paper.

The shift symmetries of natural inflation have also been used in
multiple field models.  Freese \cite{term} suggested using a PNGB as
the rolling field in double field inflation \cite{doublefield} (in
which the inflaton is a tunneling field whose nucleation rate is
controlled by its coupling to a rolling field).  Kaplan and Weiner
have examined natural inflation-like models in the context of
``little'' fields \cite{kapwei}.

Kawasaki {\it et al.} proposed a supergravity inflation model in which
the inflaton potential is flat due to a shift symmetry, again
utilizing the basic idea of natural inflation.  Choi {\it et al.} have
discussed thermal inflation in the context of radiatively generated
axion scale in supersymmetric axion models \cite{choi97}.

While arguments based on theoretical prejudice are useful guidelines
for model building, the ultimate test is an observational one. It is a
remarkable coincidence that the borderline between consistent and
inconsistent effective field theory models for inflation is roughly
the same as the borderline between whether or not tensor modes are, in
a practical sense, observable \cite{kinney2003}. In order for the
tensor/scalar ratio to be large enough to be measured by foreseeable
future experiments, the width of the potential must be of order $f
\sim m_{\rm Pl}$ or larger. Natural inflation is still very much in
the running as a candidate model for the early universe.  These
models are also especially attractive from a particle physics
perspective because they possess a hierarchy of scales which is stable
against radiative corrections. Such a hierarchy is necessary for the
suppression of density perturbations, but, in other models,
 is typically left to an unexplained fine
tuning of the inflaton self-coupling to order $\lambda \sim 10^{-14}$.
In Natural Inflation this heirarchy of scales arises naturally.
Finally, the simplest models of natural inflation predict a large
enough gravitational wave component that a detection by advanced CMB
measurements will be possible.

\section*{Acknowledgments}

K.F. thanks Tom Banks for helpful conversations.
K. F. acknowledges support from the DOE as well as the Michigan
Center for Theoretical Physics via the University of Michigan for support.
WHK thanks the Perimeter Institute and the MCTP, where part of this work
was completed, for hospitality.

\end{document}